\documentstyle[aps]{revtex}

\begin{document}
\title{Lorentz-Covariant Quantization of Massive Non-Abelian Gauge Fields in The
Hamiltonian Path-Integral Formalism }
\author{Jun-Chen Su}
\address{Center for Theoretical Physics, Department of Physics, Jilin University,\\
Changchun 130023, People's Republic of China}
\date{}
\maketitle

\begin{abstract}
The massive non-Abelian gauge fields are quantized Lorentz-covariantly in
the Hamiltonian path-integral formalism. In the quantization, the Lorentz
condition, as a necessary constraint, is introduced initially and
incorporated into the massive Yang-Mills Lagrangian by the Lagrange
multiplier method so as to make each temporal component of a vector
potential to have a canonically conjugate counterpart. The result of this
quantization is confirmed by the quantization performed in the Lagrangian
path-integral formalism by applying the Lagrange multiplier method which is
shown to be equivalent to the Faddeev-Popov approach.
\end{abstract}

\section{Introduction}

It was shown in [1,2] that the massive non-Abelian gauge fields can well be
quantized in the Hamiltonian path-integral formalism following the procedure
initiated in [3,4] for the massless gauge fields. In this quantization, the
massive Yang-Mills Lagrangian density written below was chosen to be the
starting point. 
\begin{equation}
{\cal L}=-\frac 14F^{a\mu \nu }F_{\mu \nu }^a+\frac 12m^2A^{a\mu }A_\mu ^a 
\eqnum{1.1}
\end{equation}
where $A_\mu ^a$ are the vector potentials for a massive gauge field, 
\begin{equation}
F_{\mu \nu }^a=\partial _\mu A_\nu ^a-\partial _\nu A_\mu ^a+gf^{abc}A_\mu
^bA_\nu ^c  \eqnum{1.2}
\end{equation}
are the field strengths and m is the mass of gauge bosons. Working with the
Lagrangian in equation (1.1), the massive non-Abelian gauge fields appear to
be second class systems. In the quantization, it is necessary to introduce a
primary constraint of second class defined by 
\begin{equation}
\Pi _0^a(x)=\frac{\partial {\cal L}}{\partial \dot A_0^a}=0  \eqnum{1.3}
\end{equation}
which was incorporated into the Lagrangian (1.1) by the Lagrange multiplier
method. The result of the quantization amounts to the one given in so-called
unitary gauge. An alternative quantization within the Hamiltonian
path-integral formalism was subsequently performed in [5] by the approach
proposed initially in [6]. In this approach, the second class non-Abelian
system is converted to a first class system by enlarging the phase space
with introducing a series of extra fields. The results of this quantization
can be compared with those obtained early in the Lagrangian path-integral
formalism by using the gauge-invariant St\"uckelberg Lagrangian [7-11]. Due
to the introduction of the extra field variables which have a correspondence
with the St\"uckelberg scalar functions, the quantized result looks much
complicated.

In this paper, a Lorentz-covariant quantization of the massive non-Abelian
gauge fields will be carried out within the framework of Hamiltonian
path-integral along the line described in our previous paper for the
quantization of massless non-Abelian gauge fields [12]. The essential point
of this quantization which is different from the previous is that the
Lorentz condition

\begin{equation}
\varphi ^a\equiv \partial ^\mu A_\mu ^a=0  \eqnum{1.4}
\end{equation}
, as a necessary constraint, is introduced from the beginning and imposed on
the massive Yang-Mills Lagrangian. This treatment is consistent with the
fact that a massive gauge field has only three polarization states and can
completely be described by the Lorentz-covariant (four-dimensionally)
transverse part of the vector potential, $A_T^{a\mu }(x)$. Whereas, the
Lorentz-covariant longitudinal part of the vector potential, $A_L^{a\mu }$,
appears to be a redundant unphysical variable which must be constrained by
introducing the Lorentz condition which implies $A_L^{a\mu }=0$.
Conventionally, the Lorentz condition is viewed as a consequence of the
field equations of motion [13] 
\begin{equation}
\partial ^\mu F_{\mu \nu }^a+m^2A_\nu ^a=j_\nu ^a  \eqnum{1.5}
\end{equation}
where $j_\mu ^a$ is the current generated by the gauge field itself. The
argument of this viewpoint is as follows. When we take divergence of the
both sides of equation (1.5) and notice the current conservation, it is
found that 
\begin{equation}
m^2\partial ^\mu A_\mu ^a=0  \eqnum{1.6}
\end{equation}
Since $m\neq 0$, the above equation leads to the Lorentz condition. This
seems to imply that the Lorentz condition has already been included in the
massive Yang-Mills Lagrangian. If so, when the Lagrangian is written in the
first order form, we should see a term in the Lagrangian which is given by
incorporating the Lorentz condition by the Lagrange multiplier method.
Nevertheless, as will be seen in the next section, there is no such a term
to appear in the Lagrangian. Therefore, the viewpoint stated above is not
reasonable. The correct procedure is to treat the Lorentz condition as a
primary constraint imposed on the massive Yang-Mills Lagrangian. In this
case, due to the Lorentz condition introduced, equation (1.6), as a trivial
identity, naturally holds, exhibiting the self-consistency of the theory.
This procedure coincides with the aforementioned procedure of the
quantization performed in [1,2,5] where the condition in equation (1.2), as
a primary constraint, is necessarily introduced and is also consistent with
the conventional canonical quantization. In the latter quantization, to
derive the free propagator of the gauge boson, one only uses the transverse
free field operator [13] 
\begin{equation}
{\bf A}_\mu ^c(x)=\int \frac{d^3k}{(2\pi )^3\sqrt{2\omega (k)}}\sum_{\lambda
=1}^3\epsilon _\mu ^\lambda (k)[{\bf a}_\lambda ^c(k)e^{-ikx}+{\bf a}%
_\lambda ^{c+}(k)e^{ikx}]  \eqnum{1.7}
\end{equation}
where $\epsilon _\mu ^\lambda (k)$ are the polarization vectors satisfying
the transversality condition 
\begin{equation}
k^\mu \epsilon _\mu ^\lambda (k)=0  \eqnum{1.8}
\end{equation}
which follows directly from the Lorentz condition.

We have a great interest to note that in the physical subspace defined by
the Lorentz condition, i.e., spanned by the transverse vector potential $%
A_T^{a\mu },$ the dynamics of massive gauge fields is gauge-invariant in
contrast to that the massive Yang-Mills Lagrangian itself is not
gauge-invariant. In fact, when we make an infinitesimal gauge
transformation[13 ]:

\begin{equation}
\delta A_\mu ^a=D_\mu ^{ab}\theta ^b  \eqnum{1.9}
\end{equation}
where 
\begin{equation}
D_\mu ^{ab}=\delta ^{ab}\partial _\mu -gf^{abc}A_\mu ^c  \eqnum{1.10}
\end{equation}
to the action given by the Lagrangian in equation (1.1), noticing the
identity $f^{abc}A^{a\mu }A_\mu ^b=0$ and applying the Lorentz condition, it
can be found that 
\begin{equation}
\delta S=-m^2\int d^4x\theta ^a\partial ^\mu A_\mu ^a=0  \eqnum{1.11}
\end{equation}
Alternatively, the gauge-invariance may also be formulated by means of the
Lagrangian represented in terms of the transverse fields 
\begin{equation}
{\cal L}=-\frac 14F_T^{a\mu \nu }F_{T\mu \nu }^a+\frac 12m^2A_T^{a\mu
}A_{T\mu }^a  \eqnum{1.12}
\end{equation}
which is obtained from equation (1.1) by applying the solution of equation
(1.4): $A_L^{a\mu }=0$. Under the gauge transformation taking place in the
physical subspace 
\begin{equation}
\delta A_{T\mu }^a=D_{T\mu }^{ab}\theta ^b  \eqnum{1.13}
\end{equation}
where $D_{T\mu }^{ab}$ is defined as that in equation (1.10) with $A_\mu ^c$
being replaced by $A_{T\mu }^c,$ it is easy to prove that the action given
by the Lagrangian (1.12) is invariant 
\begin{equation}
\delta S=-m^2\int d^4x\theta ^a\partial ^\mu A_{T\mu }^a=0  \eqnum{1.14}
\end{equation}
where the transversality condition: $\partial ^\mu A_{T\mu }^a=0,$ as an
identity, has been noticed. The gauge-invariance of the action in the
physical subspace suggests that the quantum non-Abelian gauge field theory
may also be set up on the basis of gauge-invariance principle as will be
specified in section 4.

The remainder of this paper is arranged as follows. In section 2, we will
start from the Lagrangian written in the first order formulation. This
Lagrangian is given by incorporating the Lorentz condition into the massive
Yang-Mills Lagrangian by the Lagrange undetermined multiplier method and
will be used to derive canonical equations of motion for the fields. It will
be demonstrated that the equations of motion derived are self-consistent and
complete for describing the field dynamics. In section3, the quantization of
the massive gauge field will be Lorentz-covariantly performed in the
Hamiltonian path-integral formalism. The result of this quantization is
different from those given in [1,2,5]. In section 4, for comparison, the
quantization will be alternatively carried out in the Lagrangian
path-integral formalism by applying the Lagrangian multiplier method and the
gauge-invariance principle. It will be shown that the Lagrangian multiplier
method is equivalent to the Faddeev-Popov approach [14]. The last section
serves to make conclusions and comments on the problems of renormalizability
and unitarity of the theory.

\section{First order formulation and equations of motion}

According to the general procedure, the Lorentz condition (1.4) may be
incorporated into the Lagrangian (1.1) by the Lagrange undetermined
multiplier method to give a generalized Lagrangian. In the first order
formalism, this Lagrangian is written as [3,4] 
\begin{equation}
{\cal L}=\frac 14F^{a\mu \nu }F_{\mu \nu }^a-\frac 12F^{a\mu \nu }(\partial
_\mu A_\nu ^a-\partial _\nu A_\mu ^a+gf^{abc}A_\mu ^bA_\nu ^c)+\frac 12%
m^2A^{a\mu }A_\mu ^a+\lambda ^a\partial ^\mu A_\mu ^a  \eqnum{2.1}
\end{equation}
where $A_\mu ^a$ and $F_{\mu \nu }^a$ are now treated as the mutually
independent variables and $\lambda ^a$ are chosen to represent the Lagrange
multipliers. Using the canonically conjugate variables defined by 
\begin{equation}
\Pi _\mu ^a(x)=\frac{\partial {\cal L}}{\partial \dot A^{a\mu }}=F_{\mu
0}^a+\lambda ^a\delta _{\mu 0}={\cal \{} 
\begin{tabular}{l}
$F_{k0}^a=E_k^a,$ if $\mu =k=1,2,3;$ \\ 
$\lambda ^a=-E_0^a,$ if $\mu =0.$%
\end{tabular}
\eqnum{2.2}
\end{equation}
the Lagrangian (2.1) may be rewritten in the canonical form 
\begin{equation}
{\cal L}=E^{a\mu }\dot A_\mu ^a+A_0^aC^a-E_0^a\varphi ^a-{\cal H} 
\eqnum{2.3}
\end{equation}
where 
\begin{equation}
C^a=\partial ^\mu E_\mu ^a+gf^{abc}A_k^bE^{ck}+m^2A_0^a  \eqnum{2.4}
\end{equation}
\begin{equation}
{\cal H}=\frac 12(E_k^a)^2+\frac 14(F_{ij}^a)^2+\frac 12%
m^2[(A_0^a)^2+(A_k^a)^2]  \eqnum{2.5}
\end{equation}
here $E_\mu ^a=(E_0^a,E_k^a)$ is a Lorentz vector, ${\cal H}$ is the
Hamiltonian density in which $F_{ij}^a$ are defined by 
\begin{equation}
F_{ij}^a=\partial _iA_j^a-\partial _jA_i^a+gf^{abc}A_i^bA_j^c  \eqnum{2.6}
\end{equation}
In the above, the four-dimensional and the spatial indices are respectively
denoted by the Greek and Latin letters. From the stationary condition of the
action constructed by the Lagrangian (2.3), one may derive the equations of
motion as follows 
\begin{equation}
\dot A_k^a=\partial _kA_0^a+gf^{abc}A_k^bA_0^c-E_k^a  \eqnum{2.7}
\end{equation}
\begin{equation}
\dot E_k^a=\partial
^iF_{ik}^a+gf^{abc}(E_k^bA_0^c+F_{ki}^bA^{ci})+m^2A_k^a+\partial _kE_0^a 
\eqnum{2.8}
\end{equation}
\begin{equation}
C^a(x)\equiv \partial ^\mu E_\mu ^a+gf^{abc}A_k^bE^{ck}+m^2A_0^a=0 
\eqnum{2.9}
\end{equation}
and equation (1.4) where $k=1,2,3$. Equations (2.7) and (2.8) act as the
equations of motion satisfied by the independent canonical variables $A_k^a$
and $E_k^a(k=1,2.3)$ which precisely describe the three degrees of freedom
of polarization for a massive gauge field with a given group index, while,
equations (1.4) and (2.9) can only be regarded as the constraint equations
obeyed by the constrained variables $A_0^a$ and $E_0^a$ because in these
equations, there are no time-derivatives of the dynamical variables $A_k^a$
and $E_k^a$. Equation (2.3) clearly shows that these constraints have
already been incorporated in the Lagrangian by the Lagrange undetermined
multiplier method. Especially, the Lagrange multipliers are just the
constrained variables themselves in this case. It is clear to see that in
equations (2.7)-(2.9) and (1.4) there are altogether eight equations for a
given group index. They are sufficient to determine the eight variables
including three pairs of the dynamical canonical variables $A_k^a$ and $%
E_k^a(k=1,2.3)$ and one pair constrained variables $A_0^a$ and $E_0^a$,
showing the completeness of the equations.

Along the general line by Dirac [3], we shall examine the evolution of the
constraints $\varphi ^a$ and $C^a$ with time. Taking the derivative of the
both equations (1.4) and (2.9) with respect to time and making use of the
equations of motion : 
\begin{equation}
\dot A_\mu ^a(x)=\frac{\delta H}{\delta E^{a\mu }(x)}-\int d^4y[A_0^b(y)%
\frac{\delta C^b(y)}{\delta E^{a\mu }(x)}-E_0^b(y)\frac{\delta \varphi ^b(y)%
}{\delta E^{a\mu }(x)}]  \eqnum{2.10}
\end{equation}
\begin{equation}
\dot E_\mu ^a(x)=-\frac{\delta H}{\delta A^{a\mu }(x)}+\int d^4y[A_0^b(y)%
\frac{\delta C^b(y)}{\delta A^{a\mu }(x)}-E_0^b(y)\frac{\delta \varphi ^b(y)%
}{\delta A^{a\mu }(x)}]  \eqnum{2.11}
\end{equation}
which are obtained from the stationary condition of the action given by the
Lagrangian (2.3), one may derive the following consistency equations [2,3] 
\begin{equation}
\{H,\varphi ^a(x)\}+\int d^4y[\{\varphi ^a(x),C^b(y)\}A_0^b(y)-\{\varphi
^a(x),\varphi ^b(y)\}E_0^b(y)]=0  \eqnum{2.12}
\end{equation}
\begin{equation}
\{H,C^a(x)\}+\int d^4y[\{C^a(x),C^b(y)\}A_0^b(y)-\{C^a(x),\varphi
^b(y)\}E_0^b(y)]=0  \eqnum{2.13}
\end{equation}
where equations (1.4) and (2.9) have been used. In the above, $H$ is the
Hamiltonian defined by an integral of the Hamiltonian density shown in
equation (2.5) over the coordinate x and $\{F,G\}$ represents the Poisson
bracket which is Lorentz-covariantly defined as 
\begin{equation}
\{F,G\}=\int d^4x\{\frac{\delta F}{\delta A_\mu ^a(x)}\frac{\delta G}{\delta
E^{a\mu }(x)}-\frac{\delta F}{\delta E_\mu ^a(x)}\frac{\delta G}{\delta
A^{a\mu }(x)}\}  \eqnum{2.14}
\end{equation}
The Poisson brackets in equations (2.12) and (2.13) are easily calculated.
The results are 
\begin{equation}
\{C^a(x),\varphi ^b(y)\}=D_\mu ^{ab}(x)\partial _x^\mu \delta ^4(x-y) 
\eqnum{2.15}
\end{equation}
\begin{equation}
\{\varphi ^a(x),\varphi ^b(y)\}=0  \eqnum{2.16}
\end{equation}
\begin{equation}
\{C^a(x),C^b(y)\}=m^2[gf^{abc}A_0^c(x)-2\delta ^{ab}\partial _0^x]\delta
^4(x-y)  \eqnum{2.17}
\end{equation}
\begin{equation}
\{H,\varphi ^a(x)\}=\partial _x^kE_k^a(x)  \eqnum{2.18}
\end{equation}
\begin{equation}
\{H,C^a(x)\}=m^2[\partial _0^xA_0^a(x)+\partial _k^xA_k^a(x)]  \eqnum{2.19}
\end{equation}
where $D_\mu ^{ab}(x)$ was defined in equation (1.10). It is pointed out
that by the requirement of Lorentz-covariance, in the computation of the
above brackets, the second term in equation (2.9) has been written in a
Lorentz-covariant form $gf^{abc}A^{b\mu }E_\mu ^c$. We are allowed to do it
because the added term $gf^{abc}A_0^bE_0^c$ only gives a vanishing
contribution to the term $A_0^aC^a$ in equation (2.3) due to the identity $%
f^{abc}A_0^aA_0^b=0$. Particularly, the determinant of the matrix which is
constructed by the Poisson bracket denoted in equation (2.15) is not
singular. This indicates that equations (2.12) and (2.13) are solvable to
determine the Lagrange multipliers $A_0^a(x)$ and $E_0^a(x)$. There is no
necessity of taking other subsidiary constraint conditions into account
further. This reveals the consistency of the constraints (1.4) and (2.9). On
substituting equations (2.15)-(2.19) into equations (2.12) and (2.13), we
find 
\begin{equation}
\Box _xA_0^a(x)-gf^{abc}\partial _x^\mu [A_0^b(x)A_\mu ^c(x)]-\partial
_x^kE_k^a(x)=0  \eqnum{2.20}
\end{equation}
\begin{equation}
\lbrack \delta ^{ab}\Box _x-gf^{abc}A_\mu ^c(x)\partial _x^\mu ]E_0^b(x)=0 
\eqnum{2.21}
\end{equation}
These equations are compatible with the equations (1.4) and (2.7)-(2.9). In
fact, as can easily be verified, equations.(2.20) and (2.21) may directly be
derived from equations (1.4) and (2.7-(2.9). In addition, we mention that if
the Hamiltonian density is defined by $\bar {{\cal H}}={\cal H}%
-A_0^aC^a+E_0^a\varphi ^a$, the equations (2.20) and (2.21) can also be
obtained from the equations $\{\bar H,\varphi ^a(x)\}=0$ and $\{\bar H%
,C^a(x)\}=0$ respectively where $\bar H$ is the Hamiltonian defined by an
integral of the $\bar {{\cal H}\text{ }}$ over the coordinate x.

\section{Quantization in the Hamiltonian path-integral formalism}

This section serves to formulate the quantization performed in the
Hamiltonian path-integral formalism for the massive non-Abelian gauge
fields. In accordance with the general procedure of the quantization [4], we
firstly write the generating functional of Green's functions via the
independent canonical variables which are now chosen to be the transverse
parts of the vectors $A_\mu ^a$ and $E_\mu ^a$ 
\begin{equation}
Z[J]=\frac 1N\int D(A_T^{a\mu },E_T^{a\mu })exp\{i\int d^4x[E_T^{a\mu }\dot A%
_{T\mu }^a-{\cal H}^{*}(A_T^{a\mu },E_T^{a\mu })+J_T^{a\mu }A_{T\mu }^a]\} 
\eqnum{3.1}
\end{equation}
where ${\cal H}^{*}(A_T^{a\mu },E_T^{a\mu })$ is the Hamiltonian which is
obtained from the Hamiltonian (2.5) by replacing the constrained variables $%
A_L^{a\mu }$ and $E_L^{a\mu }$ with the solutions of equations (1.4) and
(2.9). As mentioned before, equation (1.4) leads to $A_L^{a\mu }=0$.
Noticing this solution and the decomposition $E^{a\mu }(x)=E_T^{a\mu
}+E_L^{a\mu }(x),$ when setting $E_L^{a\mu }(x)=\partial _x^\mu Q^a(x)$
where $Q^a(x)$ is a scalar function, one may get from equation (2.9) an
equation obeyed by the scalar function $Q^a(x)$ for a given group index 
\begin{equation}
K^{ab}(x)Q^b(x)=W^a(x)  \eqnum{3.2}
\end{equation}
where 
\begin{equation}
K^{ab}(x)=\delta ^{ab}\Box _x-gf^{abc}A_T^{c\mu }(x)\partial _\mu ^x 
\eqnum{3.3}
\end{equation}
and 
\begin{equation}
W^a(x)=gf^{abc}E_T^{b\mu }(x)A_{T\mu }^c(x)-m^2A_T^{a0}(x)  \eqnum{3.4}
\end{equation}
With the aid of the Green's function $G^{ab}(x-y)$ (the ghost particle
propagator) which satisfies the following equation 
\begin{equation}
K^{ac}(x)G^{cb}(x-y)=\delta ^{ab}\delta ^4(x-y)  \eqnum{3.5}
\end{equation}
one may find the solution to the equation (3.2) as follows 
\begin{equation}
Q^a(x)=\int d^4yG^{ab}(x-y)W^b(y)  \eqnum{3.6}
\end{equation}
With the expressions given in equations (3.4) and (3.6), we see that the $%
E_L^{a\mu }(x)$ is a complicated functional of the variables $A_T^{a\mu }$
and $E_T^{a\mu }$ so that the Hamiltonian ${\cal H}^{*}(A_T^{a\mu
},E_T^{a\mu })$ is of much more complicated functional structure which is
not convenient for constructing the diagram technique in perturbation
theory. Therefore, it is better to express the generating functional in
equation (3.1) in terms of the variables $A_\mu ^a$ and $E_\mu ^a$. For this
purpose, it is necessary to insert the following delta-functional into
equation (3.1) 
\begin{equation}
\delta [A_L^{a\mu }]\delta [E_L^{a\mu }-E_L^{a\mu }(A_T^{a\mu },E_T^{a\mu
})]=detM\delta [C^a]\delta [\varphi ^a]  \eqnum{3.7}
\end{equation}
where $M$ is the matrix whose elements are 
\begin{equation}
M^{ab}(x,y)=\{C^a(x),\varphi ^b(y)\}  \eqnum{3.8}
\end{equation}
which was given in equation (2.15). The relation in equation (3.7) is easily
derived from equations (1.4) and (2.9) by applying the property of
delta-functional [4]. Upon inserting equation (3.7) into equation (3.1) and
utilizing the Fourier representation of the delta-functional 
\begin{equation}
\delta [C^a]=\int D(\eta ^a/2\pi )e^{i\int d^4x\eta ^aC^a}  \eqnum{3.9}
\end{equation}
we have 
\begin{eqnarray}
Z[J] &=&\frac 1N\int D(A_\mu ^a,E_\mu ^a,\eta ^a)detM\delta [\partial ^\mu
A_\mu ^a]exp\{i\int d^4x[E^{a\mu }\dot A_\mu ^a  \nonumber \\
&&\ +\eta ^aC^a-{\cal H}(A_\mu ^a,E_\mu ^a)+J^{a\mu }A_\mu ^a]\} 
\eqnum{3.10}
\end{eqnarray}
In the above exponential, there is a $E_0^a$-related term $E_0^a(\partial
_0A_0^a-\partial _0\eta ^a)$ which permits us to perform the integration
over $E_0^a$, giving a delta-functional $\delta [\partial _0A_0^a-\partial
_0\eta ^a]=det|\partial _0|^{-1}\delta [A_0^a-\eta ^a]$. The determinant $%
det|\partial _0|^{-1}$, as a constant, may be put in the normalization
constant $N$ and the delta-functional $\delta [A_0^a-\eta ^a]$ will
disappear when the integration over $\eta ^a$ is carried out. The integral
over $E_k^a$ is of Gaussian-type and hence easily calculated. After these
manipulations, we arrive at 
\begin{eqnarray}
Z[J] &=&\frac 1N\int D(A_\mu ^a)detM\delta [\partial ^\mu A_\mu
^a]exp\{i\int d^4x[-\frac 14F^{a\mu \nu }F_{\mu \nu }^a  \nonumber \\
&&\ +\frac 12m^2A^{a\mu }A_\mu ^a+J^{a\mu }A_\mu ^a]\}  \eqnum{3.11}
\end{eqnarray}
When employing the familiar expression [4] 
\begin{equation}
detM=\int D(\bar C^a,C^a)e^{i\int d^4xd^4y\bar C^a(x)M^{ab}(x,y)C^b(y)} 
\eqnum{3.12}
\end{equation}
where $\bar C^a(x)$ and $C^a(x)$ are the mutually conjugate ghost field
variables and the following limit for the Fresnel functional 
\begin{equation}
\delta [\partial ^\mu A_\mu ^a]=\lim_{\alpha \to 0}C[\alpha ]e^{-\frac i{%
2\alpha }\int d^4x(\partial ^\mu A_\mu ^a)^2}  \eqnum{3.13}
\end{equation}
where $C[\alpha ]\sim \prod_x(\frac i{2\pi \alpha })^{1/2}$ and
supplementing the external source terms for the ghost fields, the generating
functional in equation (3.11) is finally given in the form 
\begin{equation}
Z[J,\overline{\xi },\xi ]=\frac 1N\int D(A_\mu ^a,\bar C^a,C^a)exp\{i\int
d^4x[{\cal L}_{eff}+J^{a\mu }A_\mu ^a+\overline{\xi }^aC^a+\bar C^a\xi ^a]\}
\eqnum{3.14}
\end{equation}
where 
\begin{equation}
{\cal L}_{eff}=-\frac 14F^{a\mu \nu }F_{\mu \nu }^a+\frac 12m^2A^{a\mu
}A_\mu ^a-\frac 1{2\alpha }(\partial ^\mu A_\mu ^a)^2-\partial ^\mu \bar C%
^aD_\mu ^{ab}C^b  \eqnum{3.15}
\end{equation}
which is the effective Lagrangian for the quantized massive gauge field in
which the first two terms are the Yang-Mills Lagrangian , the third and
fourth terms are the so-called gauge-fixing term and the ghost term
respectively. In equation (3.14), the limit $\alpha \to 0$ is implied.
Certainly, the theory may be given in general gauges $(\alpha \ne 0)$. In
this case, as will be seen in the next section, the ghost particle will
acquire a spurious mass $\mu =\sqrt{\alpha }m$.

\section{Quantization in the Lagrangian path-integral formalism}

To help understanding of the result of the quantization given in the
preceding section, in this section, we attempt to quantize the massive
non-Abelian gauge fields in the Lagrangian path-integral formalism. For
later convenience, the massive Yang-Mills Lagrangian in equation (1.1) and
the Lorentz constraint condition in equation (1.4) are respectively
generalized to the following forms

\begin{equation}
{\cal L}_\lambda =-\frac 14F^{a\mu \nu }F_{\mu \nu }^a+\frac 12m^2A^{a\mu
}A_\mu ^a-\frac 12\alpha (\lambda ^a)^2  \eqnum{4.1}
\end{equation}
and

\begin{equation}
\partial ^\mu A_\mu ^a+\alpha \lambda ^a=0  \eqnum{4.2}
\end{equation}
where $\lambda ^a(x)$ are the extra functions which will be identified with
the Lagrange multipliers and $\alpha $ is an arbitrary constant playing the
role of gauge parameter. Now, according to the general procedure for
constrained systems, equation (4.2) may be incorporated into equation (4.1)
by the Lagrange multiplier method, giving a generalized Lagrangian

\begin{equation}
{\cal L}_\lambda =-\frac 14F^{a\mu \nu }F_{\mu \nu }^a+\frac 12m^2A^{a\mu
}A_\mu ^a+\lambda ^a\partial ^\mu A_\mu ^a+\frac 12\alpha (\lambda ^a)^2 
\eqnum{4.3}
\end{equation}
This Lagrangian is obviously not gauge-invariant. However, for building up a
correct gauge field theory, it is necessary to require the dynamics of the
gauge field to be gauge-invariant. In other words, the action given by the
Lagrangian (4.3) is required to be invariant under the gauge transformations
shown in equation (1.9). By this requirement, noticing the identity $%
f^{abc}A^{a\mu }A_\mu ^b=0$ and applying the constraint condition (4.2), we
have

\begin{equation}
\delta S_\lambda =-\frac 1\alpha \int d^4x\partial ^\nu A_\nu ^a(x)\partial
^\mu ({\cal D}_\mu ^{ab}(x)\theta ^b(x))=0  \eqnum{4.4}
\end{equation}
where 
\begin{equation}
{\cal D}_\mu ^{ab}(x)=\delta ^{ab}\frac{\mu ^2}{\Box _x}\partial _\mu
^x+D_\mu ^{ab}(x)  \eqnum{4.5}
\end{equation}
in which $\mu ^2=\alpha m^2$ and $D_\mu ^{ab}(x)$ was defined in equation
(1.10). From equation (4.2) we see $\frac 1\alpha \partial ^\nu A_\nu
^a=-\lambda ^a\ne 0$. Therefore, to ensure the action to be gauge-invariant,
the following constraint condition on the gauge group is necessary to be
required 
\begin{equation}
\partial _x^\mu ({\cal D}_\mu ^{ab}(x)\theta ^b(x))=0  \eqnum{4.6}
\end{equation}
These are the coupled equations satisfied by the parametric functions $%
\theta ^a(x)$ of the gauge group. Since the Jacobian is not singular 
\begin{equation}
detM\ne 0  \eqnum{4.7}
\end{equation}
where 
\begin{eqnarray}
M^{ab}(x,y) &=&\frac{\delta (\partial _x^\mu {\cal D}_\mu ^{ac}(x)\theta
^c(x))}{\delta \theta ^b(y)}  \nonumber \\
&=&\delta ^{ab}(\Box _x+\mu ^2)\delta ^4(x-y)-gf^{abc}\partial _x^\mu (A_\mu
^c(x)\delta ^4(x-y))  \eqnum{4.8}
\end{eqnarray}
the above equations are solvable and would give a set of solutions which
express the functions $\theta ^a(x)$ as functionals of the vector potentials 
$A_\mu ^a(x)$. The constraint conditions in equation (4.6) may also be
inserted into the Lagrangian (4.3) by the Lagrange undetermined multiplier
method. In doing this, it is convenient, as usually done, to introduce ghost
field variables $C^a(x)$ in such a fashion [13] 
\begin{equation}
\theta ^a(x)=\zeta C^a(x)  \eqnum{4.9}
\end{equation}
where $\zeta $ is an infinitesimal Grassmann's number. In accordance with
equation (4.9), the constraint condition (4.6) can be rewritten as 
\begin{equation}
\partial ^\mu ({\cal D}_\mu ^{ab}C^b)=0  \eqnum{4.10}
\end{equation}
where the number $\zeta $ has been dropped. This constraint condition
usually is called ghost equation. When the condition (4.10) is incorporated
into the Lagrangian (4.3) by the Lagrange multiplier method, we obtain a
more generalized Lagrangian as follows 
\begin{equation}
{\cal L}_\lambda =-\frac 14F^{a\mu \nu }F_{\mu \nu }^a+\frac 12m^2A^{a\mu
}A_\mu ^a+\lambda ^a\partial ^\mu A_\mu ^a+\frac 12\alpha (\lambda ^a)^2+%
\bar C^a\partial ^\mu ({\cal D}_\mu ^{ab}C^b)  \eqnum{4.11}
\end{equation}
where $\bar C^a(x)$, acting as Lagrange undetermined multipliers, are the
new scalar variables conjugate to the ghost variables $C^a(x).$

At present, we are ready to formulate the quantization of the massive gauge
field . As we learn from the Lagrange undetermined multiplier method, the
dynamical and constrained variables as well as the Lagrange multipliers in
the Lagrangian (4.11) can all be treated as free ones, varying arbitrarily.
Therefore, we are allowed to use this kind of Lagrangian to construct the
generating functional of Green's functions 
\begin{eqnarray}
Z[J^{a\mu },\overline{\xi }^a,\xi ^a] &=&\frac 1N\int D(A_\mu ^a,\bar C%
^a,C^a,\lambda ^a)exp\{i\int d^4x[{\cal L}_\lambda (x)+J^{a\mu }(x)A_\mu
^a(x)  \nonumber \\
&&\ \ +\overline{\xi }^a(x)C^a(x)+\bar C^a(x)\xi ^a(x)]\}  \eqnum{4.12}
\end{eqnarray}
where $D(A_\mu ^a,\cdots ,\lambda ^a)$ denotes the functional integration
measure, $J_\mu ^a,\bar K^a$ and $K^a$ are the external sources coupled to
the gauge and ghost fields and $N$ is a normalization constant. Looking at
the expression of the Lagrangian (4.11), we see, the integral over $\lambda
^a(x)$ is of Gaussian-type. Upon completing the calculation of this
integral, we finally arrive at 
\begin{eqnarray}
Z[J^{a\mu },\overline{\xi }^a,\xi ^a] &=&\frac 1N\int D(A_\mu ^a,\bar C%
^a,C^a,)exp\{i\int d^4x[{\cal L}_{eff}(x)  \nonumber \\
&&\ \ +J^{a\mu }(x)A_\mu ^a(x)+\overline{\xi }^a(x)C^a(x)+\bar C^a(x)\xi
^a(x)]\}  \eqnum{4.13}
\end{eqnarray}
where 
\begin{equation}
{\cal L}_{eff}=-\frac 14F^{a\mu \nu }F_{\mu \nu }^a+\frac 12m^2A^{a\mu
}A_\mu ^a-\frac 1{2\alpha }(\partial ^\mu A_\mu ^a)^2-\partial ^\mu \bar C^a%
{\cal D}_\mu ^{ab}C^b  \eqnum{4.14}
\end{equation}
is the effective Lagrangian given in the general gauges. In the Landau gauge
($\alpha \rightarrow 0$), The Lagrangian (4.14) goes over to the one given
in equation (3.15). When the mass m tends to zero, equation (4.14) is
immediately converted to the Lagrangian encountered in the massless gauge
field theory.

Now let us turn to show that the quantization described above is equivalent
to the quantization by the Faddeev-Popov approach. According to the
procedure of the latter approach, we need to insert the following identity
[14] 
\begin{equation}
\Delta [A]\int D(\theta ^a)\delta [\partial ^\mu A_\mu ^\theta +\alpha
\lambda ^\theta ]=1  \eqnum{4.15}
\end{equation}
into the vacuum-to-vacuum transition amplitude, obtaining 
\begin{equation}
Z[0]\equiv <0^{+}\mid 0^{-}>=\frac 1N\int D(A_\mu ^a,\lambda ^a,\theta
^a)\Delta [A]\delta [\partial ^\mu A_\mu ^\theta +\alpha \lambda ^\theta
]exp(iS_\lambda )  \eqnum{4.16}
\end{equation}
where $S_\lambda $ is the action given by the Lagrangian (4.1)

\begin{equation}
S_\lambda =\int d^4x[-\frac 14F^{a\mu \nu }F_{\mu \nu }^a+\frac 12m^2A^{a\mu
}A_\mu ^a-\frac 12\alpha (\lambda ^a)^2]  \eqnum{4.17}
\end{equation}
The delta-functional in equation (4.16) implies

\begin{equation}
\partial ^\mu (A^\theta )_\mu ^a+\alpha (\lambda ^\theta )^a=\partial ^\mu
A_\mu ^a+\alpha \lambda ^a=0  \eqnum{4.18}
\end{equation}
where $(A^\theta )_\mu ^a$ and $(\lambda ^\theta )^a$ represent the
gauge-transformed vector potentials and Lagrange multipliers, $(A^\theta
)_\mu ^a=A_\mu ^a+\delta A_\mu ^a$ here $\delta A_\mu ^a$ was denoted in
equation (1.9) and $(\lambda ^\theta )^a=\lambda ^a+\delta \lambda ^a$.
Equation (4.18) holds naturally because the constraint condition (4.2) is
required to be satisfied for all the field variables including the ones
before and after gauge transformation. The $\delta \lambda ^a$ may be
determined by the requirement that the action $S_\lambda $ is to be
gauge-invariant with respect to the gauge transformations of $A_\mu ^a$ and $%
\lambda ^a$. By this requirement, it is easy to find

\begin{equation}
\delta S_\lambda =\int d^4x\partial ^\mu A_\mu ^a(\delta \lambda
^a-m^2\theta ^a)=0  \eqnum{4.19}
\end{equation}
where the condition in equation (4.2) has been considered. From the above
equation, noticing $\partial ^\mu A_\mu ^a\neq 0$ in the general gauges, we
see, it must be 
\begin{equation}
{\delta \lambda }^a{=m^2\theta }^a.  \eqnum{4.20}
\end{equation}
Here we see that in the case of massless gauge fields, ${\delta \lambda }%
^a=0 $ so that equation (4.19) becomes a trivial identity. When the gauge
transformations given in equations (1.9) and (4.20) are inserted into
equation (4.18), we may obtain a constraint condition which is identical to
that denoted in equation (4.6). Hence, from equation (4.15) we get $\Delta
[A]=\det $M[A] in which the matrix M[A] is completely the same as given in
equation (4.8). It is easy to verify that the determinant det M[A] , the
integration measure and the action in equation (4.17) are all invariant with
respect to the gauge transformations of the functions $A_\mu ^a$ and $%
\lambda ^a.$ Therefore, when we make the gauge transformations: $A_\mu
^a\rightarrow (A^{-\theta })_\mu ^a$ and $\lambda ^a\rightarrow (\lambda
^{-\theta })^a$ to the functional integral in equation (4.16), the integral
over $\theta ^a(x)$ , as a constant, may be factored out from the functional
integral over $A_\mu ^a$ and $\lambda ^a$ and put in the normalization
constant N. Thus, we have

\begin{equation}
Z[0]=\frac 1N\int D(A_\mu ^a,\lambda ^a)\det M[A]\delta [\partial ^\mu A_\mu
+\alpha \lambda ]exp(iS_\lambda )  \eqnum{4.21}
\end{equation}
On completing the integration over $\lambda ^a$, we obtain 
\begin{equation}
Z[0]=\frac 1N\int D(A_\mu ^a)detM[A]exp\{iS-\frac i{2\alpha }\int
d^4x(\partial ^\mu A_\mu ^a)^2\}  \eqnum{4.22}
\end{equation}
where S is the massive Yang-Mills action. By making use of the
representation of det M[A] as shown in equation (3.12) and introducing the
external sources, we exactly recover the generating functional written in
equations (4.13) and (4.14).

In the end, we would like to emphasize that under the Lorentz condition
which has been introduced into the Lagrangian\ in equation (4.3) and the
transition amplitude in equation (4.16), the gauge-invariance of the action
shown in equation (4.4) and equation (4.19) was formulated only for the
infinitesimal gauge transformations denoted in equation (1.9). The quantized
result shown in equations (4.13) and (4.14) was derived just by the use of
such gauge transformations. In the Landau gauge ($\alpha =0)$, this result
is exactly the same as that obtained in section 3 by the quantization
performed in the Hamiltonian path-integral formalism for which we only need
to calculate the classical Poisson brackets without concerning any gauge
transformation. This fact indicates that to get the correct quantized result
by the methods formulated in this section, the infinitesimal gauge
transformations are only necessary to be taken into account. This implies
that in the physical subspace restricted by the Lorentz condition, only the
infinitesimal gauge transformations are possible to exist.

\section{Conclusions and comments}

The new features of this paper consist in: (1) The Lorentz-covariant
quantization is successfully achieved in the Hamiltonian path-integral
formalism; (2) The result of this quantization is confirmed by the
quantization performed by the Lagrange multiplier method in the Lagrangian
path-integral formalism. The latter method is proposed first in this paper
and shown to be equivalent to the Faddeev-Popov approach; (3) The
established quantum massive non-Abelian gauge field theory can be
straightforwardly converted to the massless theory in the zero-mass limit in
contrast to the previous quantization in the Hamiltonian formulation [1,2,5]
for which the zero-mass limit does not exist. The essential points to
achieve these results are clarified in this paper. They include: (1) The
Lorentz condition, as a necessary constraint, must be introduced initially
and imposed on the massive Yang-Mills Lagrangian so as to give a complete
formulation of the dynamics of massive gauge fields. That is to say, in the
whole space of the full vector potential, the massive non-Abelian gauge
fields must be considered as constrained systems and the massive Yang-Mills
Lagrangian itself is not complete for describing the dynamics; (2) The
Lorentz condition may be incorporated into the massive Yang-Mills Lagrangian
by the Lagrange multiplier method so that each component of a vector
potential acquires its canonically conjugate counterpart. This makes the
Lorentz-covariant formulation of the Hamiltonian path-integral quantization
become possible; (3) In the physical subspace restricted by the Lorentz
condition, the action other than the Lagrangian for the massive gauge field
is gauge-invariant. Since the action is of more fundamental dynamical
meaning than the Lagrangian, the gauge-invariance of the action implies that
the dynamics of the massive gauge field is gauge-invariant. Therefore , the
quantum massive non-Abelian gauge field theory may also be set up on the
basis of gauge-invariance; (4) In the physical subspace, only infinitesimal
gauge transformations are possibly allowed. This point explicated clearly in
the end of the former section was already pointed out in reference [14]. In
the reference, after introducing the identity denoted in equation (4.15)
into the vacuum-to vacuum transition amplitude, the authors said that '' We
must know $\Delta [A]$ is only for the transverse fields and in this case
all contributions to the last integral are given in the neighborhood of the
unite element of the group''. By this point, it can be understood why in the
ordinary quantum gauge field theories such as the standard model, the
BRST-transformations are all taken to be infinitesimal.

Now we are in a position to comment on the previous works for the massive
gauge field theories. Originally, the theory was set up from the massive
Yang-Mills Lagrangian alone which is viewed as complete for describing the
field dynamics [15-17]. To overcome the gauge-non-invariance of the mass
term in the Lagrangian, the St\"uckelberg Lagrangian in which the mass term
is written in a gauge-invariant form by introducing the extra St\"uckelberg
field functions was proposed and widely chosen to be the starting point to
establish the quantum theory [7-11]. All these theories were considered to
be nonrenormalizable due to the appearance of the unphysical gauge degrees
of freedom in the theories which arises from the finite gauge
transformations. We would like to point out that according to the basic
ideas stated before, the Lorentz condition should be , as a primiry
constraint, introduced initially into the theories and thereby only the
infinitesimal gauge transformations are necessary to be taken into account.
Based on this point, let us analysize the previous works mentioned above. In
reference [15], the authors proved an equivalence theorem by which they gave
a Hamiltonian derived from the massive Yang-Mills Lagrangian by introducing
an auxiliary St\"uckelberg field functions. When making an unitary
transformation to the Schr\"odinger equation they derived, the mass term in
the Hamiltonian becomes dependent on the auxiliary field and contains an
infinite number of terms in its expansion of power series which lead to bad
unrenormalizability. In reference [16], the equivalence theorem was given in
the form of S-matrix. The author also introduced the St\"uckelberg field and
used it to make a finite gauge transformation to the fields involved in the
theory. As a result, the mass term in the S-matrix contains an exponential
function of the auxiliary field which gives rise to an infinite variety of
distinct primitively divergent graphs that can not be eliminated by the
introduced renormalizability conditions imposed on the gauge transformation
(see equations (39) and (40) in the reference). It is noted here that when
infinitesimal gauge transformations are, as they should be, considered only
in the theory, it is easy to find that the gauge degrees of freedom are
reduced and, particularly, the introduced renormalizability conditions are
perfectly satisfied. In reference [17], the author found a relation by which
any full vector potential may be represented as a gauge transformation of
the physical transverse vector potential and tried to separate the gauge
degrees of freedom from the transverse ones as it is able to be done for the
renormalizable theory of the massive neutral gauge boson coupled to a
conserved current. He eventually failed to do it because in the non-Abelian
case the coupling between the both degrees of freedom does not vanish upon
integration and thereof he concluded that the theory is nonrenormalizable.
However, this conclusion is invalid for the infinitesimal gauge
transformations required by the Lorentz condition because in this case the
coupling mentioned above vanishes and, therefore, the gauge degree of
freedom can also be separated out. As a result, the generating functional,
as in the case of massive neutral gauge boson theory, may also be expressed
through the transverse fields like equation (3.1). In reference [13], the
quantization of the massive non-Abelian gauge field was implemented by the
Faddeev-Popov operation with incorporation of the Lorentz condition in the
generating functional. Nevertheless, the incorporation of the Lorentz
condition is only for the purpose of improving the behavior of the massive
gauge boson propagator, being not thought of a necessary procedure. In
addition, different from the procedure stated in section 4, the quantization
does not meet the requirement of gauge-invariance so that the ghost particle
loses a mass term in the general gauges ($\alpha \neq 0$). In particular,
opposite to the procedure that the ghost term in the effective Lagrangian is
introduced by applying the infinitesimal gauge transformations, the finite
gauge transformations are made to the effective Lagrangian. As a result of
these transformations, the mass term depends on the parametric functions of
the gauge group and contains various unrenormalizable infinities. As pointed
out before, when the Lorentz condition is incorporated in the theory, a
consistent procedure is that the infinitesimal gauge transformations are
required only. In this case, the unrenormalizable infinities in the mass
term could not appear. Let us turn to the gauge-invariant St\"uckelberg
Lagrangian [8-10]

\begin{equation}
L=-\frac 14F^{a\mu \nu }F_{\mu \nu }^a+\frac 12m^2(A_\mu ^a-\omega _\mu
^a)(A^{a\mu }-\omega ^{a\mu })  \eqnum{5.1}
\end{equation}
where $\omega _\mu ^a$ are the group-valued St\"uckelberg functions defined
as 
\begin{equation}
\omega _\mu ^a=\frac ig(\partial _\mu U^{-1}U)^a=(\frac{e^{i\omega }-1}{%
i\omega })^{ab}\partial _\mu \phi ^b  \eqnum{5.2}
\end{equation}
in which $U=e^{ig\phi ^aT^a}$ and $\omega ^{ab}=igf^{abc}\phi ^c.$ The
function $\omega _\mu ^a$ for each group index a contains an infinite number
of terms when we expand the exponential function $e^{i\omega }$ as a series.
The quantum theory built by the Lagrangian (5.1) was proved to be
nonrenormalizable due to the nonpolynomial nature of the function ${\omega }%
_\mu ^a$. It is noted that in the Lagrangian (5.1), the vector field $\omega
_\mu ^a$ is also given a mass. Therefore, if it is considered to be
physical, having three polarized states, we should also impose on it the
Lorentz condition 
\begin{equation}
\partial ^\mu \omega _\mu ^a=0  \eqnum{5.3}
\end{equation}
Substituting equation (5.2) in equation (5.3), it will be found that a
necessary and sufficient condition to meet equation (5.3) is $\partial _\mu
\phi ^a=0$ which according to equation (5.2) leads to $\omega _\mu ^a=0.$
This implies that the St\"uckelberg functions are unnecessary to be
introduced. On the other hand, if the functions $\omega _\mu ^a$ are treated
as free variables, not receiving any constraint, there should be an integral
over them in the generating functional. The integral is of Gaussian-type and
hence easily calculated. As a result of the integration, the mass term of
the gauge fields will completely be cancelled out from the Lagrangian. Thus,
we are left with only a massless gauge field theory. In a word, if the
quantization of the massive gauge fields respets the basic requirements
stated in the beginning of this section as was done in the preceding
sections, the gauge degrees of freedom will be reduced and the problem of
the nonrenormalizability will disappear.

From the perturbative expansion of the generating functional given in
equations (4,13) and (4.14), it is easy to derive the bare vertices and free
propagators. The vertices are the same as those given in the massless
theory. The gauge boson propagator and the ghost particle one are
respectively, in the momentum space, of the forms

\begin{equation}
iD_{\mu \nu }^{ab}(k)=-i\delta ^{ab}\{\frac{g_{\mu \nu }-k_\mu k_\nu /k^2}{%
k^2-m^2+i\varepsilon }+\frac{\alpha k_\mu k_\nu /k^2}{k^2-\mu
^2+i\varepsilon }\}  \eqnum{5.4}
\end{equation}
and 
\begin{equation}
i\Delta ^{ab}(q)=\frac{-i\delta ^{ab}}{q^2-\mu ^2+i\varepsilon }  \eqnum{5.5}
\end{equation}
These propagators show good renormalizable behavior in the large momentum
limit, indicating that the power counting argument is applicable in this
case to support the renormalizability of th theory. The renormalizability of
the theory originates from the fact that the unphysical degrees of freedom
contained in the massive Yang-Mills Lagrangian, i e., the longitudinal
component of the gauge field and the residual gauge degrees of freedom
existing in the physical subspace are respectively constrained by the
introduced Lorentz condition and the ghost equation. The both constraints
respectively give rise to the gauge-fixing term and the ghost term in the
effective Lagrangian (4.14) which just play the role of counteracting the
unphysical degrees of freedom in the massive Yang-Mills Lagrangian. This
cancellation also ensures the unitarity of the theory. It is easy to check
that for the theory of the gauge field coupled to a vector current such as
QCD, all the S-matrix elements given in the tree diagram approximation are
unitary because the unphysical longitudinal part of the gauge boson
propagator gives a vanishing contribution to the S-matrix elements. For the
higher order perturbative S-matrix elements, it can also be proved that the
unphysical intermediate states are cancelled out with each other so that the
unitarity of the theory is still preserved ( The proofs will be reported
later). The problem arises for the theory of the charged gauge boson coupled
to a charged chiral current as we met in the electroweak theory [13]. For
this kind of theory , it seems that the tree-unitarity is violated if
without recourse to the Higgs mechanism [18,19]. We do not concern this kind
of theory in this paper. But, we would like to note that even for the
charged gauge boson theory without introducing the Higgs mechanism, the
unitarity of the theory may also be preserved by means of the limiting
procedure proposed originally in reference [20]. The limiting procedure
means that the gauge boson propagator in equation (5.4) which is written in
the $\alpha $-gauge will, in the limit: $\alpha \rightarrow \infty $, be
converted to the one given in the unitary gauge 
\begin{equation}
iD_{\mu \nu }^{ab}(k)=-i\delta ^{ab}\frac{g_{\mu \nu }-k_\mu k_\nu /m^2}{%
k^2-m^2+i\varepsilon }  \eqnum{5.6}
\end{equation}
which was originally derived in the canonical quantization by making use of
the Fourier representation of the transverse vector potential denoted in
equation (1.7). Since the $\alpha $-gauge propagator has good renormalizable
behavior, one may employ this kind of propagator to calculate the S-matrix
and then by the limiting procedure to obtain the unitary gauge result so as
to guarantee the unitarity of the theory. In doing this, the theory given in
the $\alpha $-gauge can be viewed as a regularization of the theory given in
the unitary gauge.

\section{\bf ACKNOWLEDGMENT}

The author would like to thank professor Shi-Shu Wu for useful discussions.
This subject was supported in part by National Natural Science Foundation of
China and the Research Fund for the Doctoral Program of Higher Education.

\section{References}

[1] P. Senjanovic, Ann. Phys. (New York) 100, 227 (1976).

[2] C.Grosse-Knetter, Phys. Rev. D48, 2854 (1993).

[3] P. A. M. Dirac, Lectures on Quantum Mechanics, Yeshiva University Press,
New York (1964).

[4] L. D. Faddeev, Theor. Math. Phys. 1, 1; L. D. Faddeev and A. A. Slavnov,

Gauge Fields: Introduction to Quantum Theory, The Benjamin Cummings
Publishing Company Inc.(1980).

[5] N. Banerjee, R. Banejee and Subir Ghosh, Ann. Phys. 241, 237 (1995).

[6] I. A. Batalin and I. V. Tyutin, Nucl. Phys. B279, 514 (1987); Inter. J.
Mod. Phys. A6, 3255 (1991).

[7] E. C. G. St\"uckelberg, Helv. Phys. Acta. 30, 225 (1957).

[8] T. Kunimasa and T. Goto, Prog. Theor. Phys. 37, 452 (1967).

[9] K.Shizuya, Nucl. Phys. B87, 255 (1975); B94, 260 (1975); B121, 125
(1977).

[10] T. Fukuda ,M. Monoa , M. Takeda and K. Yokoyama, Prog. Theor. Phys. 66,
1827 (1981); 67, 1206 (1982);

70, 284 (1983).

[11] A. Burnel, Phys. Rev. D33, 2981 (1986); D33, 2985 (1986).

[12] J. C. Su, J. Phys. G: Nucl. Part. Phys. 27, 1493 (2001).

[13] C. Itzykson and F-B. Zuber, Quantum Field Theory, McGraw-Hill, New York
(1980).

[14] L. D. Faddeev and V. N. Popov, Phys. Lett. B25, 29 (1967).

[15] H. Umezawa and S. Kamefuchi, Nucl. Phys. 23, 399 (1961).

[16] A. Salam, Phys. Rev. 127, 331 (1962).

[17] D .G. Boulware, Ann. Phys. 56, 140 (1970).

[18] C. H. Llewellyn-Smith, Phys. Lett. B46, 233 (1973).

[19]J. M. Cornwall, D. N. Levin and G. Tiktopoulos, Phys. Rev. Lett. 30,
1268 (1973).

[20] T. D. Lee and C. N. Yang, Phys. Rev. 128, 885 (1962).

\end{document}